\documentclass{moriond}

\def\Journal#1#2#3#4{{#1} {\bf #2}, #3 (#4)}

\def\PRD{{\em Phys. Rev.} D}

\def\be{\begin{equation}}
\def\ee{\end{equation}}
\def\bea{\begin{eqnarray}}
\def\eea{\end{eqnarray}}

\newcommand{\Photo}{\includegraphics[height=30mm]{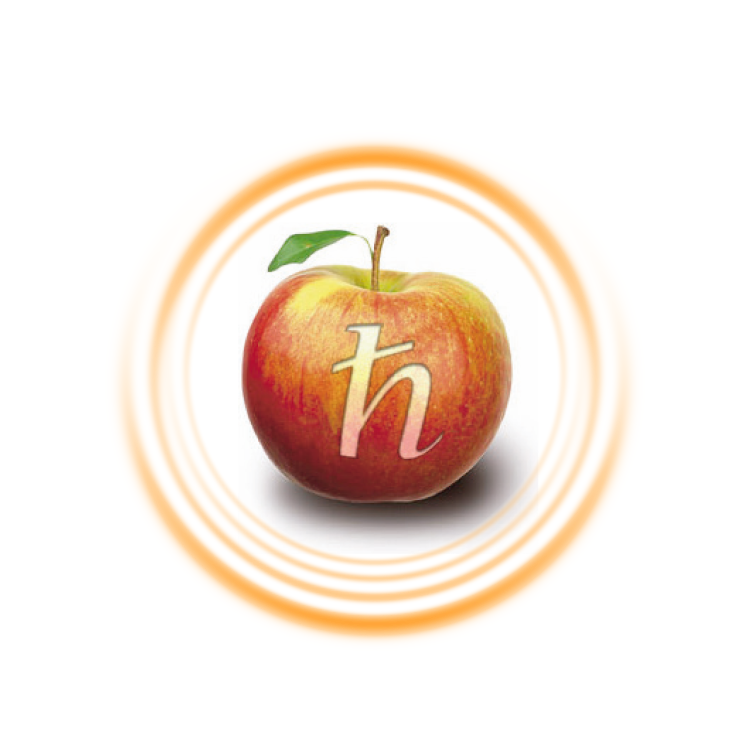}}

\begin{document}
\vspace*{4cm}
\title{QBOUNCE: FIRST MEASUREMENT OF THE NEUTRON ELECTRIC CHARGE WITH A RAMSEY-TYPE GRS EXPERIMENT}

\author{JOACHIM BOSINA$^{a,b}$ , H. FILTER$^{c}$, J. MICKO$^{a,b}$ , T. JENKE$^b$, M. PITSCHMANN$^a$, S. ROCCIA$^b$, R.I.P. SEDMIK$^a$, H. ABELE$^a$ 
}

\address{$^a$ Atominstitut, TU Wien, Austria\\
$^b$ Institut Laue-Langevin, Grenoble, France\\
$^c$ Physik Department - E66, TU München, Germany}

\maketitle\abstracts{
The \textit{q}\textsc{Bounce} collaboration built over the last years a new Ramsey-type Gravitational Resonance Spectroscopy (GRS) experiment. After commissioning between 2016 and 2018, the setup was able to measure the Ramsey transitions with GRS for the first time. Here we present a search of the hypothetical charge of the neutron as an application of GRS to study nonstandard model interactions. This article will describe the measuring principle and the setup in detail.}

\section{Introduction}

The \textit{q}\textsc{Bounce} collaboration uses ultra-cold neutrons (UCN) in order to study gravitation in the micrometer range. UCNs are ideal test probes. As all neutrons they interact electromagnetically only via their magnetic moment of their spin. Shielding the setup from magnetic fields is enough to suppress all electromagnetic disturbances. A special property of the UCN is that they hardly penetrate matter because their kinetic energy is lower than the Fermi pseudo potential of the material (e.g. aluminum: $V_F=54$\,neV). They are ideally reflected under any angle. On flat surfaces (e.g. mirrors) UCNs have to bounce off and gravity pulls them back towards the surface. Due to this confinement quantum mechanics predicts bound states which are known as the \emph{Quantum Bouncer}. The existence of these states was observed in 2002 \cite{Nesvizhevsky2002}. The last decades the techniques to manipulate these states improved a lot. Especially, the Gravitational Resonance Spectroscopy (GRS) \cite{Jenke2011-NaturePhysics-GRS} allowed to study state transitions and test different Dark Matter or Dark Energy models \cite{Cronenberg2018}.

The last major experimental step forward was the implementation of a Ramsey-type GRS setup \cite{Abele2010,Rechberger2018}. This experiment has been installed at the PF2 beam site of the Institut Laue-Langevin (ILL) since 2016. With the Ramsey-type setup we were able to search for the electric charge of the neutron as suggested in 2011 \cite{Durstberger2011}.

\begin{figure}
	\centering
	\includegraphics[width=0.9\linewidth]{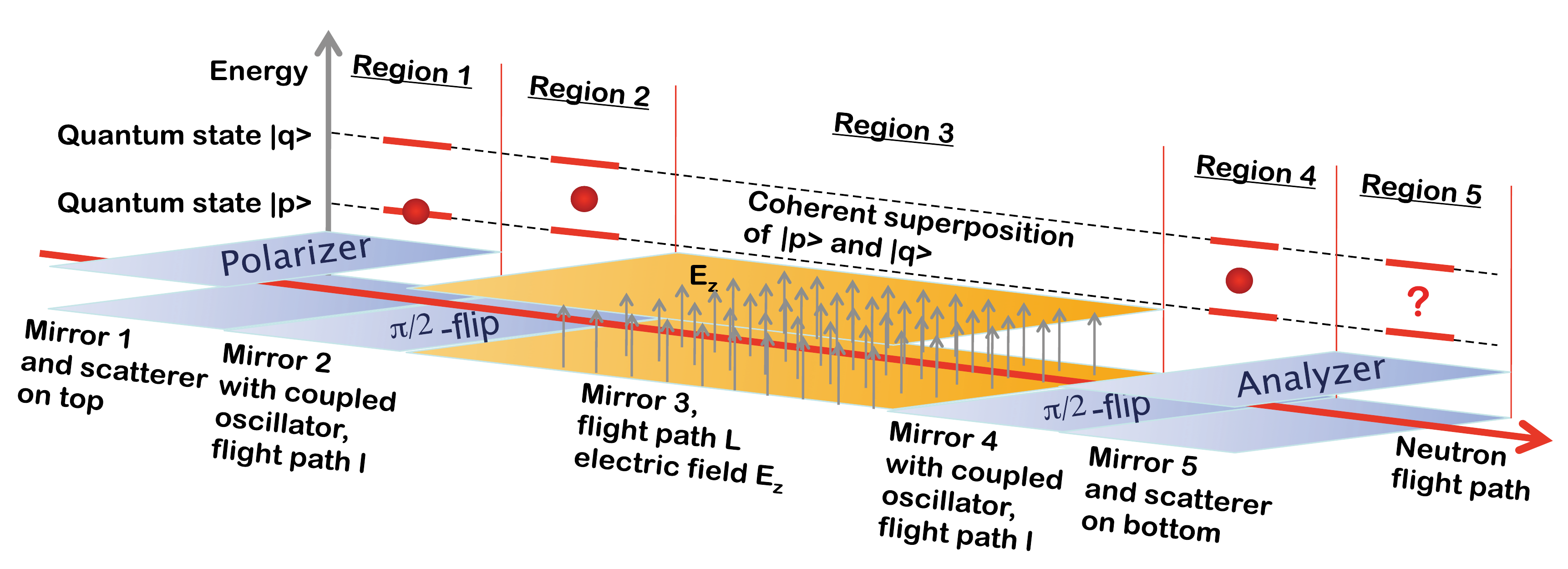}
	\caption[]{Measurement principle of the electric charge of the neutron with a Ramsey-type GRS setup (source \cite{Durstberger2011})}
	\label{fig:RamsEy}
\end{figure}
A Ramsey-type setup consists of five regions as seen in figure \ref{fig:RamsEy}. Region 1 selects only UCNs in the lowest gravitational bound states. They are further guided through the lane of mirrors to the last region. There again only the lowest states can pass through and reach the detector. Mechanical oscillation applied in region 2 and 4 can induce state transitions to higher states if the frequency and amplitude correspond to the transition energy. These higher state cannot pass region 5 and at resonance a drop in the count rate is observed. Within a Ramsey setup region 2 only excites to superposition of both states. This further evolves in region 3. For the electric charge measurement an applied electric field in region 3 induces a phase shift if a hypothetical non zero charge of the neutron exists. This would be measured as a change of the transition frequency.

\section{Theory}

\begin{figure}
	\centering
	\includegraphics[width=0.4\linewidth]{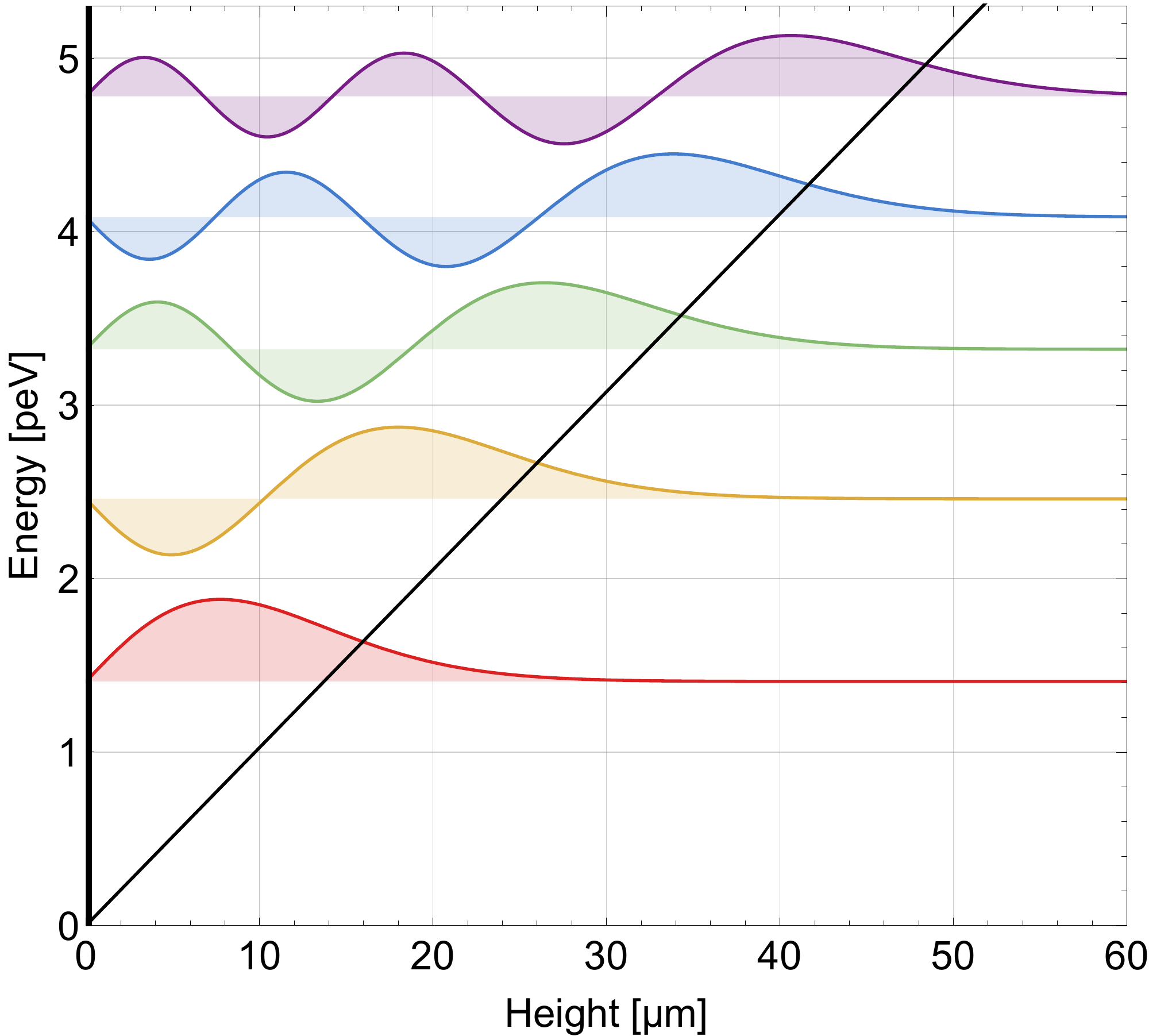}
	\caption[]{Neutron's wave functions in the linearized Newtonian gravitational potential on the surface of the Earth}
	\label{fig:wavefunction}
\end{figure}

The Schrödinger equation (Eq. \ref{eq:GraviEfieldBoundSchroedinger}) describes the motion of an ultra-cold neutron (UCN) above a flat surface (with $m_i$ and $m_g$ as the inertial and the gravitational mass, $g$ as the local gravitational acceleration). It already includes the effects of a hypothetical electric charge of the neutron $q_{\mathrm{n}}$ in an applied electric field $|\vec{E}_z|$:
\begin{equation}
	\left(-\frac{\hbar^2}{2m_i}\frac{\partial^2}{\partial z^2}+\left(m_g g\pm q_{\mathrm{n}}|\vec{E}_z|\right)z \right) \psi(z)=E \psi(z)
	\label{eq:GraviEfieldBoundSchroedinger}
\end{equation}
With a coordinate transformation $\tilde{z}=z/z_0$ and  $\tilde{E}=E/E_0$ the Schrödinger equation can be rewritten as the dimensionless Airy equation. The needed characteristic length and energy scale are $z_0=\sqrt[3]{\frac{\hbar^2}{2m_i m_g g}}$ and $E_0= m_g g z_0=\sqrt[3]{\frac{\hbar^2 m_g^2 g^2}{2m_i}}$ respectively in the case of a purely gravitational interaction. Taking an hypothetical electric charge of the neutron into account, they change to $z_0=\sqrt[3]{\frac{\hbar^2}{2m_i \left(m_g g+q_{\mathrm{n}}|\vec{E}_z|\right)}}$ and $E_0= m_g g z_0=\sqrt[3]{\frac{\hbar^2 \left(m_g g+q_{\mathrm{n}}|\vec{E}_z|\right)^2}{2m_i}}$. For a neutron on the Earth's surface the characteristic scaling parameters are $z_0=5.689\,$\textmu m and $E_0=0.602\,\mathrm{peV}$.\\
The Airy functions Ai and Bi solve the dimensionless Airy differential equation \cite{Suda2021}. Only the Airy functions Ai fulfills the boundary conditions $\psi(0)=\psi(\infty)=0$. The corresponding wave functions are parts from this function (from a root (AiZ) to infinity):
\begin{equation}
	\psi_n(z)=\frac{(-1)^{n+1}\mathrm{Ai}\left(\frac{z}{z_0}+\mathrm{AiZ}(n)\right)}{\sqrt{z_0}\mathrm{Ai}'\left(\mathrm{AiZ}(n)\right)}
	\label{eq:AirySolution1FlatSurface}
\end{equation}
As seen in figure \ref{fig:wavefunction} the wave functions are microscopically large but the energy of the bound state $E_n=-E_0\mathrm{AiZ}(n)$ is five orders of magnitude lower than the average kinetic energy of an UCN. However, neutrons with nearly all their kinetic energy in the horizontal momentum plane can populated even the lowest bound states.

The energy states are non equidistant and their differences are converted to unique frequencies which have are typically in the range of 100-1000\,Hz or higher. In order to induce a state transition we apply a mechanical oscillation to the mirror below. This correspond to introducing a time-depending periodic change of the boundary condition expressed as an additional potential $V_{GRS}(t)=a \sin(\omega t + \phi)\Theta(-z)V_{F}$ (with the Heaviside function $\Theta(z)$ as the boundary condition and $a, \omega$ and $\phi$ as the amplitude, angular frequency and the phase of the applied oscillation). This method is called Gravitational Resonance Spectroscopy (GRS).\\
Only having one oscillating region between two selector regions is a Rabi-type setup. The transmission $T_{fi}$ of a low state $i$ near a transition frequency $\omega_{fi}$ to an higher state $f$ which cannot pass the second selector can be calculated as (using the matrix element of the transition $V_{fi}=\int \psi_f (z) \partial_z \psi_i (z) \, \mathrm{d}z$, the Rabi frequency $\Omega_R=\sqrt{a^2 \omega^2 V_{fi}^2+(\omega-\omega_{fi})^2}$ and $c_{fi}$ for the state-population depending contrast):
\begin{equation}
	T_{fi}(t) = 1-c_{fi}\left(\frac{a \omega V_{fi}}{\Omega_R}\right)^2\sin^2\left(\frac{\Omega_R t}{2}\right)
	\label{eq:RabiTransistionProbability}
\end{equation}
Within a Ramsey type setup two half Rabi pulses with an intermediate free propagation region act on the neutron wave functions. This changes the transmission to the more complicated form with the advantage of a higher energy resolution \cite{Moriond2019} (with $t$ and $T$ as the flight times of a UCN through one oscillating region or region 3):
\bea
	T_{fi}(t,T)=1-c_{fi}4\sin^2\left(\frac{\Omega_R t}{2}\right) \left(\frac{a \omega V_{fi}}{\Omega_R}\right)^2
	\left( \cos\left(\frac{\Omega_R t}{2}\right) \cos\left(\frac{(\omega-\omega_{fi})T}{2}\right) \right.
	\nonumber \\
	\left.
	- \frac{(\omega-\omega_{fi})}{\Omega_R} \sin\left(\frac{\Omega_R t}{2}\right) \sin\left(\frac{(\omega-\omega_{fi})T}{2}\right)\right)
	\label{eq:RamseyTransistionProbability}
\eea
The goal of GRS is to determine the transition frequency $\omega_{fi}$ in order to check for the consistency of the underlying theory (Eq. \ref{eq:GraviEfieldBoundSchroedinger}). For the measurement of the neutron charge we also take into account the potential $V_{E_z}(z)=\pm q_{\mathrm{n}}|\vec{E}_z|z$. A none vanishing charge $q_{\mathrm{n}}$ will change the transition frequency depending on the strength of the applied electric field:
\begin{equation}
	\Delta\omega_{fi}=\omega_{fi}^0-\omega_{fi}^E=\omega_{fi}^0\left(1-\sqrt[3]{\left(1 +\frac{q_{\mathrm{n}}E_z}{m_{\mathrm{n}}g}\right)^2}\right) \approx \omega_{fi}^0\frac{2q_{\mathrm{n}}E_z}{3m_{\mathrm{n}}g}
\end{equation} Vice versa, measuring the transition frequencies at different electric field strength can determine the electric charge or at least set limits to it if no statistically significant change of the transition frequency is observed:
\begin{equation}
	q_{\mathrm{n}} \approx \frac{\Delta\omega_{fi}}{\omega_{fi}^0}\frac{3m_{\mathrm{n}}g}{2E_z}
	\hspace{30 mm}
	\delta q_{\mathrm{n}} \approx q_{\mathrm{n}} \frac{\sqrt{\left(\delta\omega_{fi}^0\right)^2+\left(\delta\omega_{fi}^E\right)^2} }{\Delta\omega_{fi}}
\end{equation}

\section{Setup}

\begin{figure}[h!]
	\centering
	\includegraphics[width=0.75\linewidth]{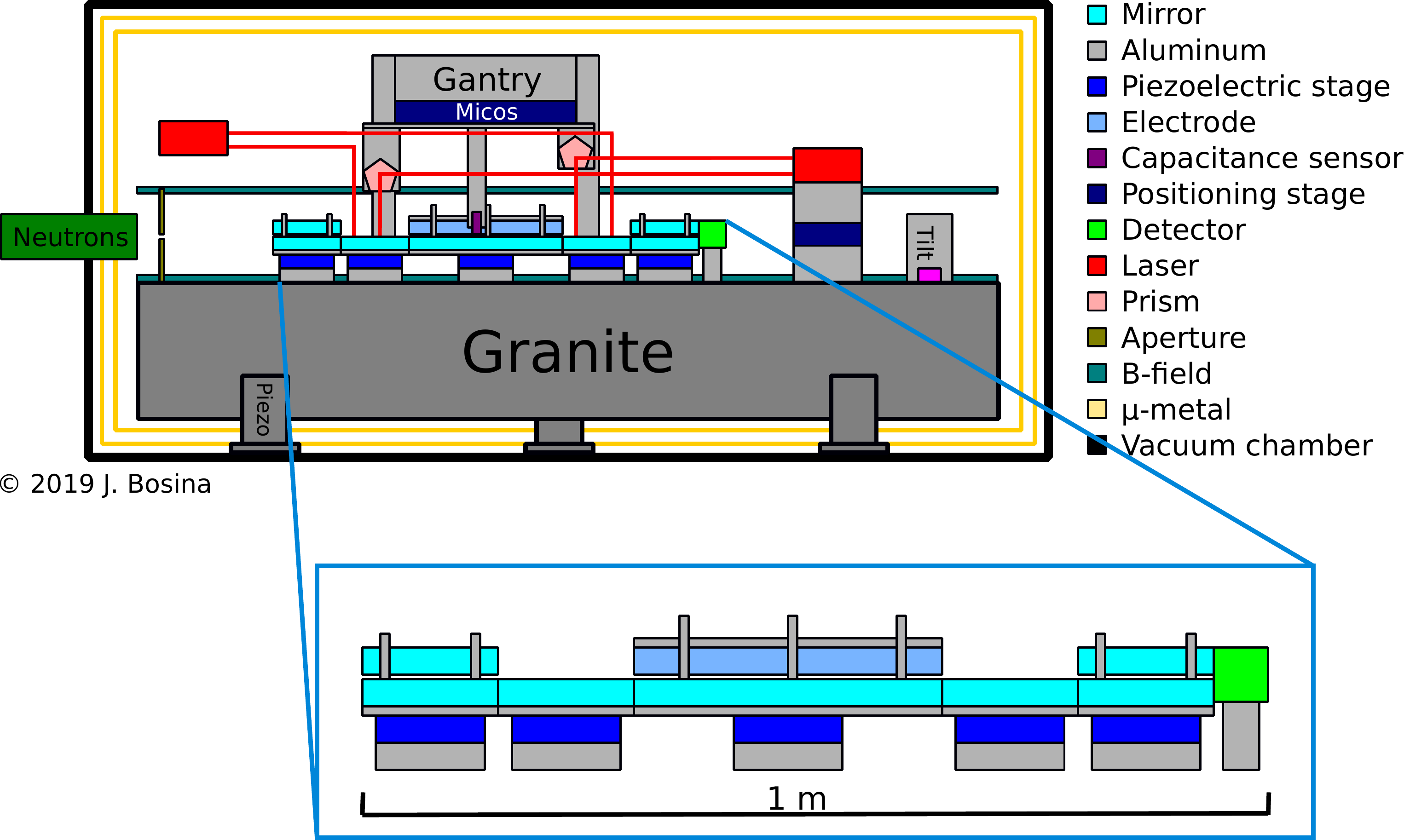}
	\caption[]{The new Ramsey-type GRS setup called Ramsey$^{\mathrm{TR}}$}
	\label{fig:SetupRamsey}
\end{figure}
As seen in figure \ref{fig:SetupRamsey} the heart of the Ramsey-type setup are five regions. Each consists of a coarse adjustment, a piezoelectric table and a mirror on top. The mirrors are made of borosilicate glass. The lengths of all regions are 152\,mm except region 3 which has a length of 340\,mm. The piezoelectric stages align the mirror surfaces down to step heights between them below 0.5\,\textmu m. An arbitrary function generator can induce mechanical oscillations to the stages of region 2 and 4. Region 1 and 5 have an additional glass plate with a rough surface clamped around 25\,\textmu m above the mirror. UCNs in a higher state have a higher probability to be close to this rough surface. Therefore, they have a high chance of either being scattered out of the system or absorbed on the surface. Only the lowest states can pass these regions and travel through the complete system. The central region 3 also acts as an electrode (parallel plate capacitor). Its mirror surface is the ground electrode and a titanium-coated second mirror is placed $\approx200$\,\textmu m above as the HV-electrode. Three fine-threaded screws connect the upper electrode to its bearing and enable to align the electrodes parallel to each other with a desired distance.\\
In order to monitor the alignment and the oscillation of the mirrors, a large gantry above the mirror lane moves different sensors. Capacitve sensors close to the mirrors monitor the alignment of the mirrors and measure the step height between the regions. The feedback of this system can be used to readjust the piezoelectric table below the mirrors in order to minimize the step height. Large prism attached to the gantry deflect beams of laser interferometers, which monitor the oscillation strength, frequency and phase, on to the mirrors.\\
All equipment is mounted on a massive granite block with a surface waviness below 2\,\textmu m within the complete area of 1900\,mm x 700\,mm. This is needed to precisely place the equipment relatively to each other. The granite also acts as a passive vibration damping stage against external disturbances. Three piezoelectric motors below the granite rock and a two-axis tilt sensor on top level the surface horizontally. The relative error of this alignment is below 1\,\textmu rad and due to analog calibrations the absolute error is 10\,\textmu rad.\\
Due to the strong absorption of UCN in air they are guided from the PF2-turbine (high flux UCN source) via evacuated steel or glass tubes to the experiment which has to be within in a large vacuum chamber. The reached vacuum pressure of $5\times10^{-5}$\,mbar is also necessary for insulating the gap between the electrodes. A \textmu -metal shielding within the vacuum chamber also suppresses external magnetic fields which would disturb the GRS measurements.\\
Between entering the vacuum chamber and region 1 a horizontal aperture selects the neutron velocity spectrum used for the measurement (5-13\,m/s). All stray neutrons are blocked and absorbed by a boron shielding housing the aperture. Only neutrons passing the aperture and all five regions can reach the detector at the end. This neutron detector is a proportional counter tube with a very low background (0.5\,mcps). Expected measurements rates are around 20\,mcps. This leads to measurement times of few hours per data point. Therefore, the setup is capable to stay at stable measurements conditions over days and weeks.

\section{Measurements}
Before and after the actual electric charge measurement of the neutron, we conducted many test with the different electrodes. The focus before was on the choice of the electrode material (copper, titanium or coated glass). The final design used aluminum coated mirrors due to their smooth surfaces and their high break down fields tested with small electrodes of around 20\,kV/mm. Additionally, they have been already used within the system for guiding the neutrons. The upper electrode was coated additionally with titanium for an electronic connections. After the neutron measurement we studied in detail the used electrode in order to find its limiting breakthrough voltage which also destroyed the coating of the electrode. The final breakthrough voltage was measured to be at 10\,kV/mm.

\begin{figure}[h!]
	\includegraphics[width=\linewidth]{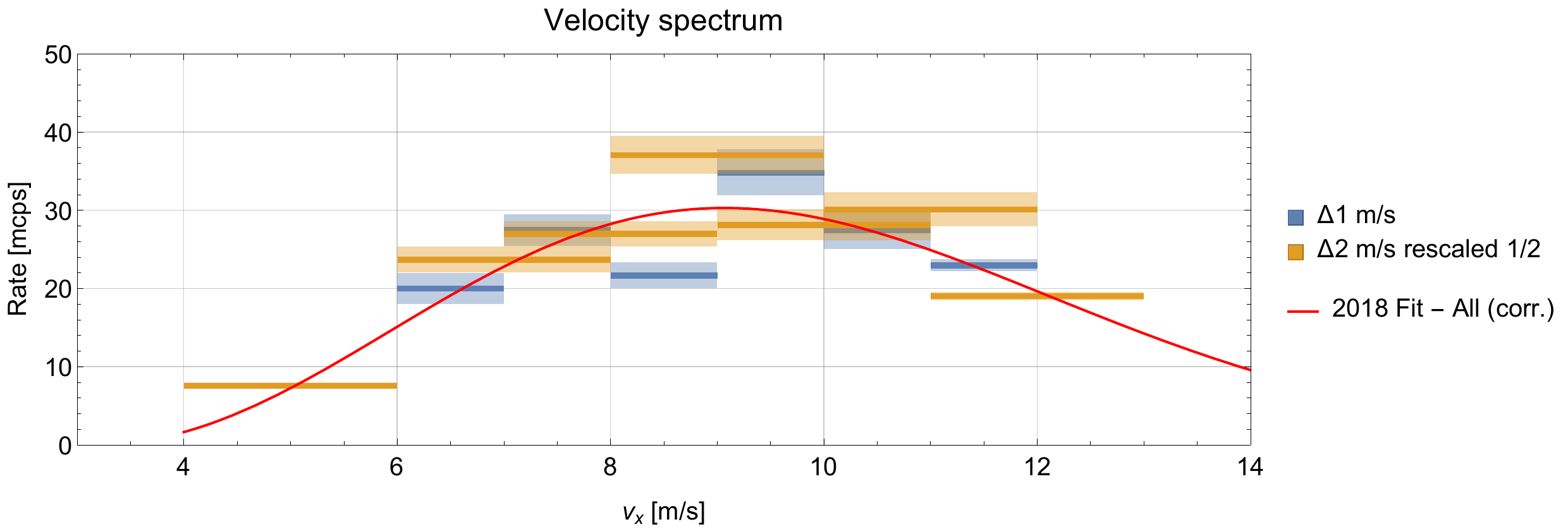}
	\caption[]{Measured velocity spectrum 2018}
	\label{fig:VelocitySpectrum}
\end{figure}

The preparatory neutron measurements were the determination of the velocity spectrum and the state population. Varying the slit setting of the aperture probed different parts of the velocity spectrum. The detector measured directly after the first region the neutron flux. The used aperture slit setting corresponded to velocity intervals of 1\,m/s (blue) or 2\,m/s (orange). The measured values can be seen in figure \ref{fig:VelocitySpectrum}. A Maxwell-Boltzmann distribution (red) was fitted to the measured spectrum. The resulting spectrum is convoluted with theory functions of Eq. \ref{eq:RabiTransistionProbability} and Eq. \ref{eq:RamseyTransistionProbability} to calculate the flight times $t$ and $T$ for neutrons with different velocities.

To determine the state selection, CR-39 detectors are placed after the first region with either region 1 or region 5 in place. These detectors have a spatial resolution below 2\,\textmu m and therefore they can image the density distributions of the wave function (Eq. \ref{eq:AirySolution1FlatSurface}). The state population of each selector region is ($47\,\%~|1\rangle$, $40\,\%~|2\rangle$, $13\,\%~|3\rangle$). The high contribution of the second state opened the possibility to drive transition from it to higher states.

A Rabi-type arrangement with only the regions 1,2 and 5 was the first attempt for GRS within the new setup and observed three transitions ($|2\rangle\rightarrow|4\rangle$, $|1\rangle\rightarrow|3\rangle$, $|2\rangle\rightarrow|5\rangle$).

\begin{figure}[h!]
	\centering
	\includegraphics[width=0.85\linewidth]{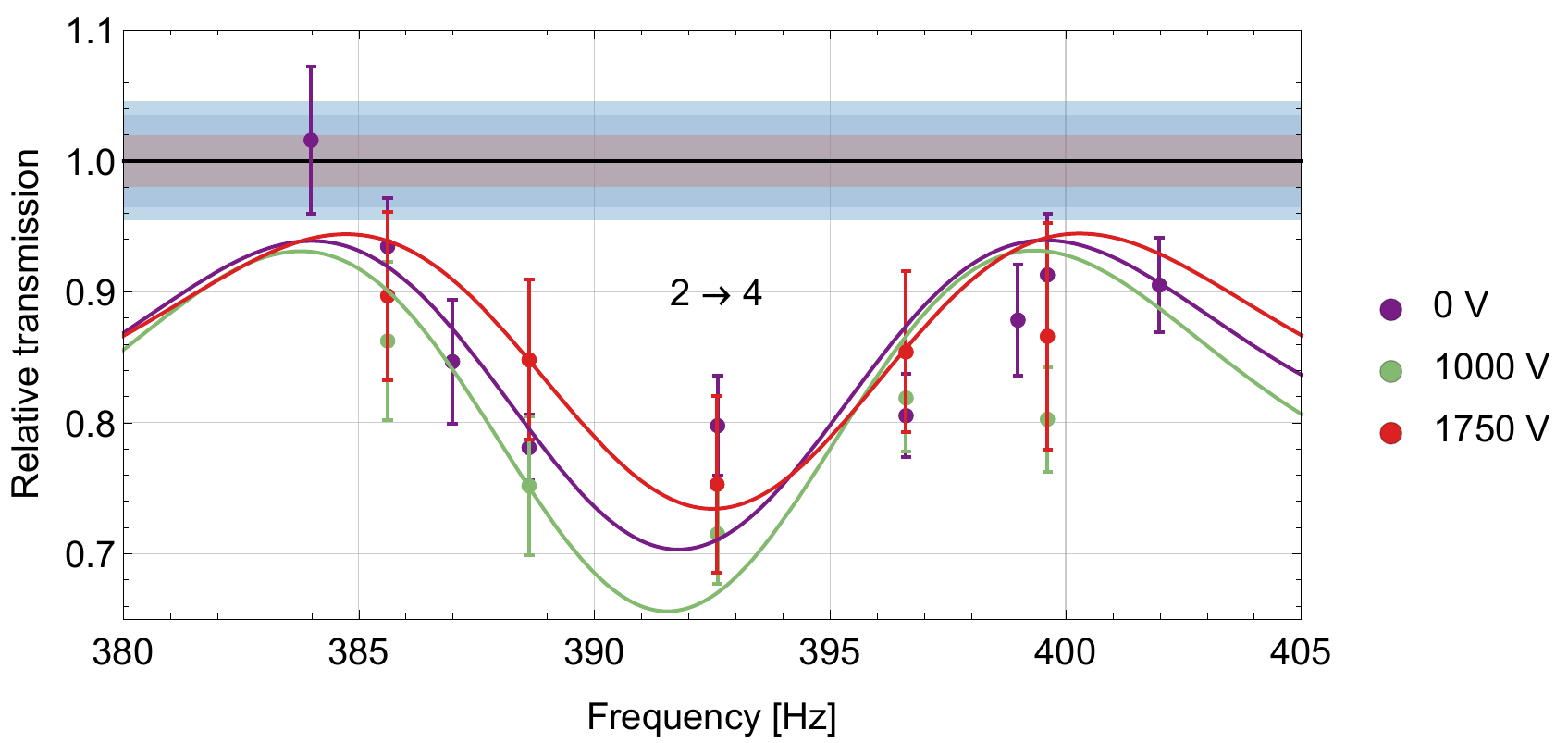}
	\caption[]{Neutron charge measurements 2018 with a Ramsey GRS setup \cite{Doi-ILL}}
	\label{fig:RamsEyPlot}
\end{figure}

The first two transition were also studied with the completed Ramsey setup. Which were the first ever measurements of this kind. The final measurements \cite{Doi-ILL} for the search for the neutron charged used only the transition $|2\rangle\rightarrow|4\rangle$ as seen in figure \ref{fig:RamsEyPlot}. The voltage settings were 1000\,V and 1750\,V at a distance of 207(1)\,\textmu m which corresponds to electric field strengths of 4.8\,kV/mm and 8.4\,kV/mm. We can estimate a statistical uncertainty of the transition frequency in the order of $\delta\omega_{fi}\sim10^{-3}$ from the counted neutrons of the measurement \cite{BosinaPhD}. This corresponds to a limit of the neutron charge in the order of $10^{-17}$ elementary charges.

\section{Conclusions}
We commissioned a new Ramsey-type GRS setup during the last years. Following the proof of principle measurements with the complete setup \cite{Rechberger2018} we used our novel approach to search for the neutron charge. The data analysis to extract statistical and systematic limits is ongoing.

The refinements with the setup since the charge measurement in 2018 suggests that the limit could be further improved with the current setup by a factor of 36: measuring bipolar (2), decreasing the gap size and increasing the field strength (2), increases in statistics ($\sqrt{10}$) and changing the transition to $|1\rangle\rightarrow|7\rangle$ (2.85). In the future, storing neutrons in the central region could improve the energy resolution enough to be able to probe the neutron charge below $10^{-21}$ elementary charges challenging the current best limit \cite{Baumann1988}.

\section*{Acknowledgments}

The \textit{q}\textsc{Bounce} experiments would not have been possible without appropriate funding of the Austrian Science Fund (FWF, project No.: P26781-N20, P33279-N, W1252). 
We are grateful for the support of the ILL, especially from the technician Thomas Brenner.

\end{document}